\documentclass[doublecol,figures]{epl2}
\pdfoutput=1
\usepackage[utf8]{inputenc}
\usepackage{graphicx}
\usepackage{amsmath}
\usepackage{amssymb}
\usepackage{gensymb} 
\usepackage{float}
\usepackage{subcaption}
\usepackage{array}
\usepackage{diagbox}
\usepackage{bm}

\title{String Shooter's overall shape in ambient air}

\author{M. Abello\inst{1,2}\thanks{E-mail: \email{margaux.abello@obspm.fr}} \and J. Courson\inst{2}\thanks{E-mail: \email{juliette.courson@orange.fr}} \and A. Maury\inst{3}\thanks{E-mail: \email{arnaud.maury@u-psud.fr}} \and J. Renaud\inst{4}\thanks{E-mail: \email{julian.renaud@ens.fr}}}
\shortauthor{M. Abello \etal}

\institute{                    
  \inst{1} Observatoire de Paris, PSL Research University - 5 Place Jules Janssen, 92190 Meudon, France\\
  \inst{2} Univ. Paris-Diderot, Université de Paris - Bâtiment Condorcet, 4 Rue Elsa Morante, 75013 Paris, France\\ 
  \inst{3} Magistère de Physique Fondamentale, Univ. Paris-Sud, Université Paris-Saclay, 91405 Orsay, France\\
  \inst{4} École Normale Supérieure, PSL Research University - 45 rue d'Ulm, 75005 Paris, France
}

\abstract{In this article, we study the behaviour of a looped string launched in ambient air using motorised wheels. We show that the loop, once it reaches its stationary state, is either in the \emph{pulley} or the \emph{air-lifted} state. The transition between these two distinct states occurs at the so-called \emph{takeoff speed}. We prove that this speed differs from one string to another based on its characteristics. However, it is independent from the loop’s length and its initial launch angle. This speed indeed corresponds to the threshold where air drag starts compensating for the weight of the string.
}

\begin{document}

\maketitle

\section{Introduction}

The \emph{String Shooter} is an experimental setup wherein a closed string loop is propelled through the air using two motorised wheels spinning in opposite directions. After some time, this loop forms a stable shape, which we call the String Shooter’s \emph{stationary state}. The concept of our setup takes inspiration from YouTube videos\cite{b.yt1,b.yt2,b.yt3,b.yt4}.

Despite its apparent simplicity, it involves several domains of mechanics, such as fluid dynamics and solid mechanics. This particularly non-intuitive behaviour led us to investigate this setup further. We had originally made a remarkable qualitative discovery in an experiment where the String Shooter was placed in a vacuum bell. The loop was first launched up in the air when at atmospheric pressure, and as we gradually decreased the air pressure, the loop fell and started dangling down, even though it was still being propelled by the spinning wheels. This made us realise the role that air drag plays in the String Shooter's physical behaviour and motivated us to study it in depth, especially since we could not find previous scientific literature on the subject.

\begin{figure}[ht]
    \centering
    \includegraphics[width=\columnwidth]{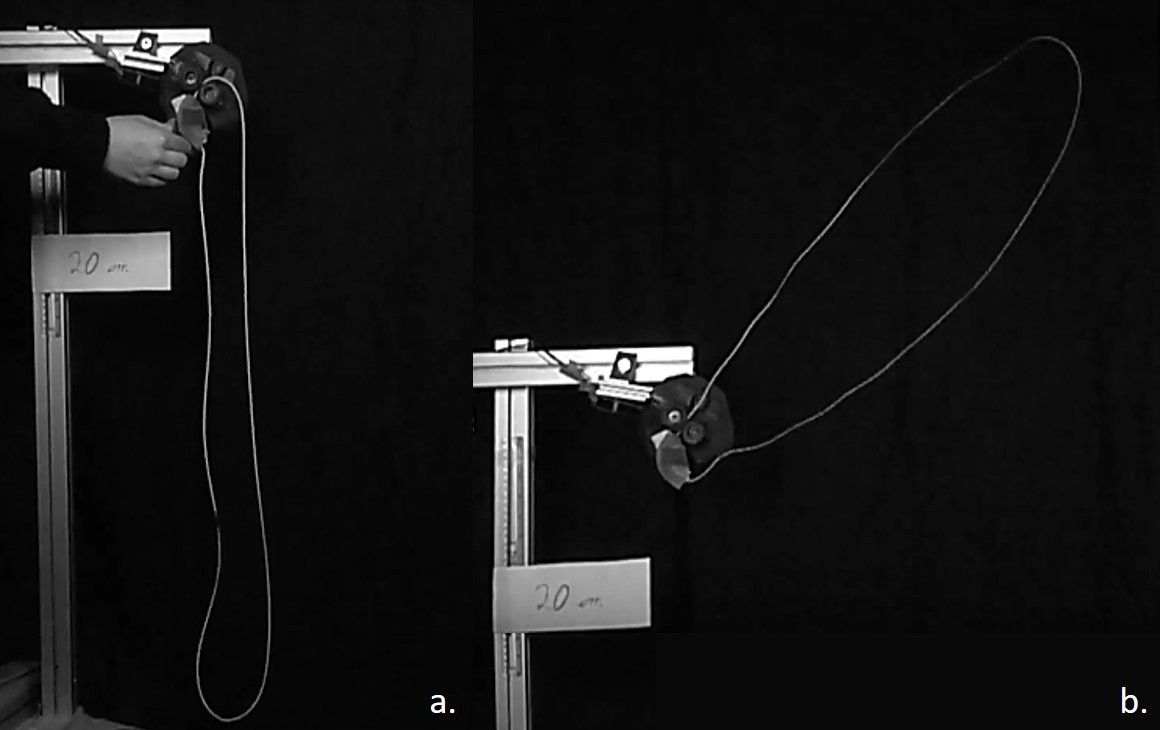}
    \caption{Photographs of the SSh setup depicting the two possible stationary states: $a)$ the pulley state; $b)$ the air-lifted state.}
    \label{fig:photoIRL}
\end{figure}

Several strings were used in this experiment, made out of three distinct materials (cotton, wool and polyester) and considered inelastic. The String Shooter also moves within a single plane, as its depth deviation is negligible compared to the string dimensions. Our goal is to study the shape of the String Shooter’s loop in this two-dimensional plane, after it reaches its stationary state (see fig~(\ref{fig:photoIRL})). For convenience, the String Shooter’s loop will simply be referred to as "\emph{SSh}".

\section{Method}

\subsection{Design of the experimental setup}

The principle of this setup consists in ejecting a piece of string in the air (at atmospheric pressure), made into a loop by making a thin knot or using strong glue, using two motorised wheels spinning at high speed. This knot or glue point forms a small imperfection in the loop which we will call the loop’s junction.

\begin{figure}[ht]
    \centering
    \includegraphics[width=\columnwidth]{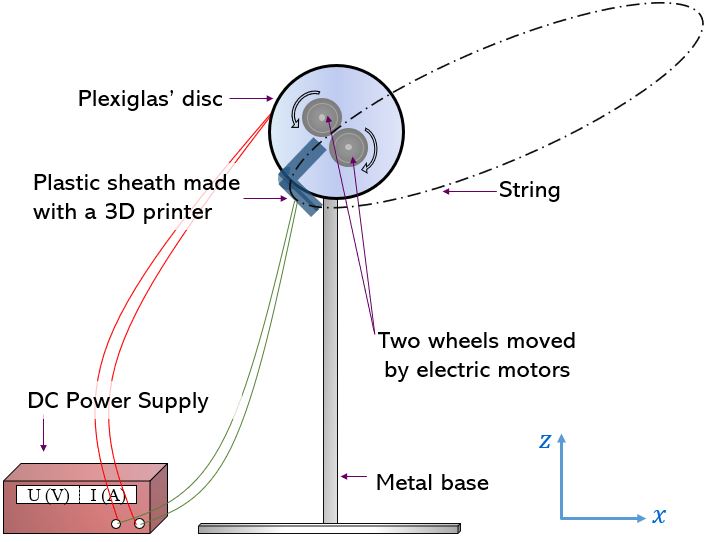}
    \caption{SSh experimental setup.}
    \label{fig:setup}
\end{figure}

\begin{figure}[ht]
    \centering
    \includegraphics[width=\columnwidth]{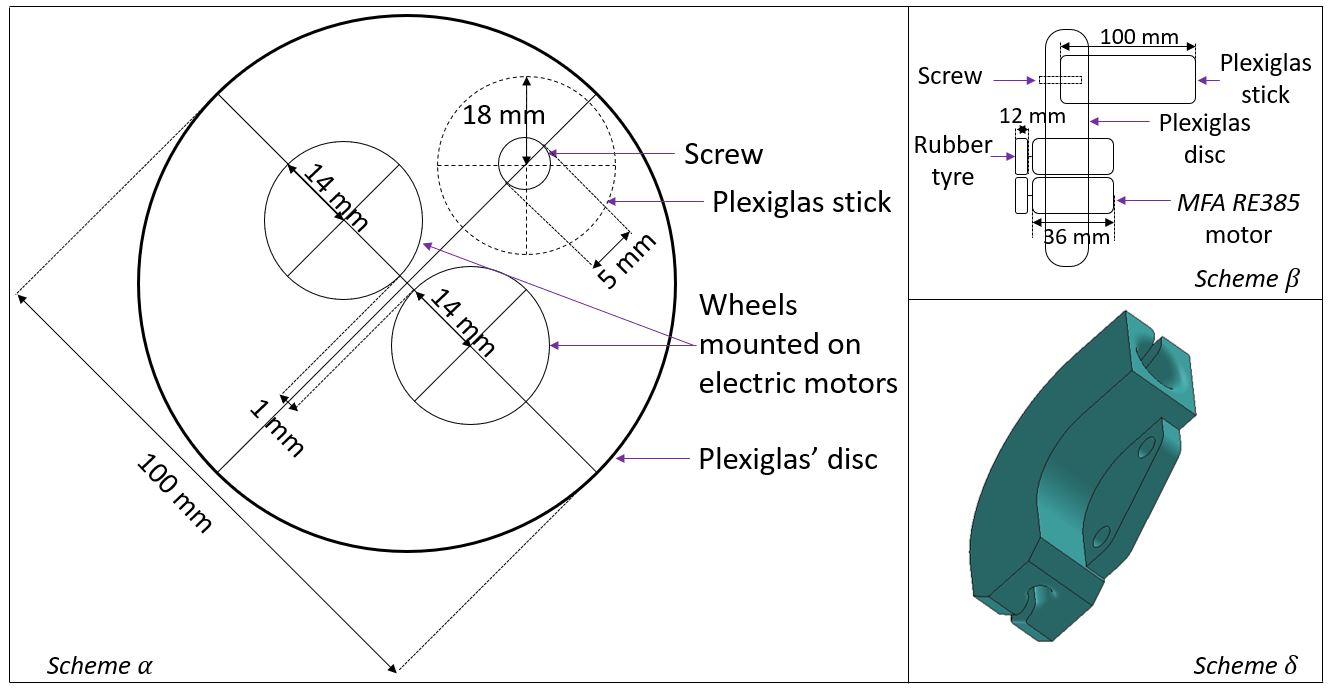}
    \caption{Schematics of the Plexiglas disk seen from the front (schematic $\alpha$), from the side (schematic $\beta$) and illustration of the sheath as designed on SolidWorks (schematic $\gamma$).}
    \label{fig:planTechnique}
\end{figure}

To build our SSh (see fig~(\ref{fig:setup})), we worked with \emph{Solarbotics RW2-3MM} wheels\footnote{found at https://www.gotronic.fr/art-roue-solarbotics-rw2-3mm-12017.htm} for their rubber tyres, ensuring an optimal grip with the string. Each wheel was mounted on a \emph{MFA RE385} motor\footnote{found at http://www.gotronic.fr/art-moteur-mfa-re385-11703.htm} (which has a maximum torque of $418\: \un{g}\cdot \un{cm}$ at 12V CC when blocked, according to the manufacturer). The two motors spin in opposite directions so as to induce a forward motion to the string, thereby ejecting it forwards. This set was then mounted and held by a Plexiglas disk. Finally, a plastic sheath (see fig~(\ref{fig:planTechnique})), crafted by a 3D printer, was placed under the wheels in such a way that it guides the returning string back in between the wheels. The exit of the sheath is located $1.7\un{cm}$ before the ejection point, where the wheels touch.

We have chosen to study several strings (see table~(\ref{tab:strings})) with different linear mass $\mu$. We have also chosen strings of different nature (cotton, wool and polyester), as for instance the polyester strings are in general smoother than wool strings on their surface.

\begin{table}
\centering
\footnotesize
\begin{tabular}{|c| c| c|}
    \hline
    \multicolumn{3}{|c|}{\textbf{Linear mass} $\bm{\mu}$ (in $\text{g}\cdot \text{m}^{-1}$)} \\
    \hline
     Polyester & Wool & Cotton \\
     \hline
     $\mathbf{p_4}$ & $\mathbf{w_2}$ & $\mathbf{c_1}$ \\
     $\mu_4 = 0.275 \pm 0.004$ & $\mu_2 = 0.221 \pm 0.004$ & $\mu_1 = 0.275 \pm 0.004$ \\ \hline
     $\mathbf{p_5}$ & $\mathbf{w_3}$ & \_\_\_\_\_\_\_\_\_\_\_\_\_\_\_\_ \\
     $\mu_5 = 0.395 \pm 0.006$ & $\mu_3 = 0.283 \pm 0.005$ &  \\ \hline
     $\mathbf{p_6}$ & \_\_\_\_\_\_\_\_\_\_\_\_\_\_\_\_ & \_\_\_\_\_\_\_\_\_\_\_\_\_\_\_\_ \\
     $\mu_6 = 2.85 \pm 0.04$ &  &  \\ \hline
     
     
\end{tabular}

\caption{List of strings used in our study, classified according to their linear mass $\mu$.}
\label{tab:strings}
\end{table}


\subsection{Tools for experimental study}

In order to study the String Shooter’s shape quantitatively, we define two unconventional but convenient distances, which we call A and B. A is defined as the distance between the ejection point (i.e. at the wheels’ intersection) and the point where the loop intercepts the tangent of slope -0.708. 
This value of the tangent's slope was chosen to intersect the slope extremity when it is in the air lifted regime. 

\begin{figure}[ht]
    \centering
    \includegraphics[width=\columnwidth]{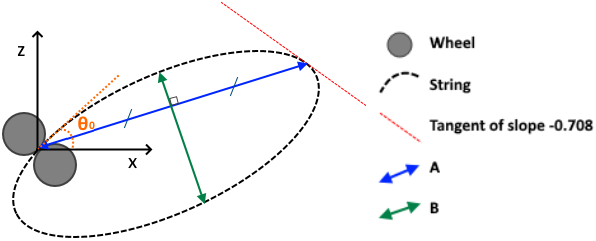}
    \caption{Definitions of the distances A and B to characterise the \emph{SSh}'s shape.}
    \label{fig:convention}
\end{figure}

Given that the analysis of this state was of particular interest, we deemed it appropriate to settle on such an accurately defined metric in our shape study. B is defined as the length between the upper branch and the lower branch of the loop along the median of A (see fig~(\ref{fig:convention})).

The SSh's physical parameters we can adjust and set are:
\begin{itemize}
    \item $v$, the speed of the loop propelled by the wheels (in $\un{m}\cdot \un{s^{-1}}$);
    \item $\mu$, the linear mass of the string (in $\un{g}\cdot \un{m^{-1}}$);
    \item $L$, the length of the string loop (in $\un{m}$);
    \item $\theta_0$, the angle of ejection of the string, formed between the horizontal axis and the tangent to the loop at the ejection point.
\end{itemize}

To measure the speed $v$ once the SSh has reached its stationary state, we record the sound made by the loop’s junction each time it passes through the sheath. Sound analysis using the Audacity \cite{b.audacity} software yields a precise measurement of the period $T_{\text{SSh}}$ of the SSh, linked to its control parameters by $v= L /T_{\text{SSh}}$.

The measurements of A, B and $\theta_0$ were made on front-view photographs of the SSh using the software ImageJ \cite{b.ImageJ} and IC Measure \cite{b.IC}.

\subsection{Theoretical model of the SSh}



At fixed velocity $v$, the SSh’s shape becomes stationary, except for some perturbations due to the imperfections of our string, our wheels, etc. These perturbations are small enough compared to the overall shape in both space and time to be neglected. Furthermore, we assume that the SSh has a negligible depth compared to its dimensions in the (x,z) plane. We shall therefore study the SSh’s stationary shape in the (x,z) plane.

We work in the laboratory’s frame of reference and parametrize the mathematical curve it forms using the arc length $s$ ranging from 0 to L. If for instance the loop’s junction is at $s=s_0$ when $t=0$, its position along the curve will be $s \equiv v \cdot t + s_0 (\text{mod L})$. This shows that it is necessary to assume the string to be inelastic, so that the length of the SSh’s loop remains constant.

\begin{figure}[ht]
    \centering
    \includegraphics[width=\columnwidth]{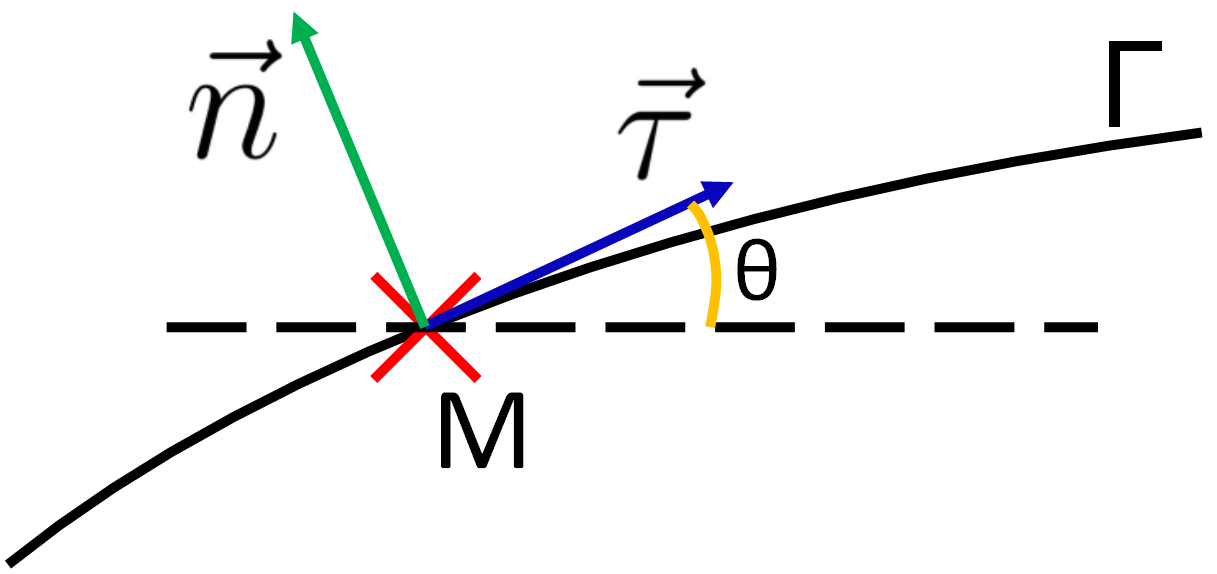}
    \caption{Illustration of the Frenet-Serret frame defined on any point M of the curve $\Gamma$.}
    \label{fig:frenet}
\end{figure}

We note $T(s)$ the tension at $s$ pulling the string towards increasing $s$.
We use the Frenet-Serret frame $(\vec{\tau}, \vec{n})$ (see  \cite{b.curv} and fig~(\ref{fig:frenet})) to describe the dynamics of an infinitesimal piece of string of length $\upd s$ subjected to its weight $\mu \upd s \vec{g}$, its tension $\frac{\upd (T \vec{\tau})}{\upd s} \upd s$ \footnote{The difference between the tensions pulling it forwards at $s+\upd s$ and pulling it backwards at $s$ is $T(s+\upd s) \vec{\tau}(s+\upd s) - T(s) \vec{\tau}(s)$.} and constant linear air drag $- f \vec{\tau}  \upd s$ with $f>0$, the requirement of such a force having been suggested by the experiment in the vacuum bell. 

\begin{figure}[ht]
    \centering
    \includegraphics[width=\columnwidth]{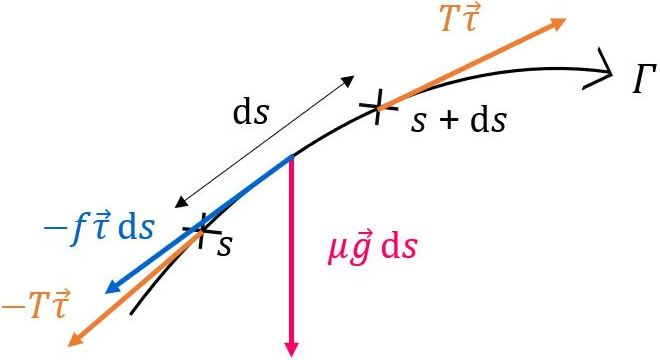}
    \caption{Illustration of all the forces acting on an infinitesimally small piece of string between $s$ and $s + \upd s$, in our model.}
    \label{fig:forces}
\end{figure}

Newton’s equation of motion yields (see fig~(\ref{fig:forces})):
\begin{equation}
\frac{\upd ( \mu v \vec{\tau} ) }{\upd t} = \mu \vec{g} + \frac{\upd (T \vec{\tau} )}{\upd s} - f \vec{\tau}.
\label{equa:Newton}
\end{equation}

Noting that $\upd s = v \cdot \upd t$, we define for convenience the effective tension $T\degree = T - \mu v ^2$ and obtain:
\begin{equation}
\frac{\upd}{\upd s} \big( T\degree \vec{\tau} \big) - f \vec{\tau} + \mu \vec{g} = \vec{0}.
\label{equa:quasiCat}
\end{equation}

Note that equation~(\ref{equa:quasiCat}) is the equation of a catenary, with a modified tension and an added tangent drag term of constant magnitude.
Recalling that $\vec{\tau} = \frac{\upd \vec{r}}{\upd s}$ and integrating with respect to s, we get:
\begin{equation}
T\degree \vec{\tau} - f \vec{r} + \mu \vec{g} s = \vec{0}.
\label{equa:vectoIntegre}
\end{equation}

\begin{figure}[ht]
    \centering
    \includegraphics[width=\columnwidth]{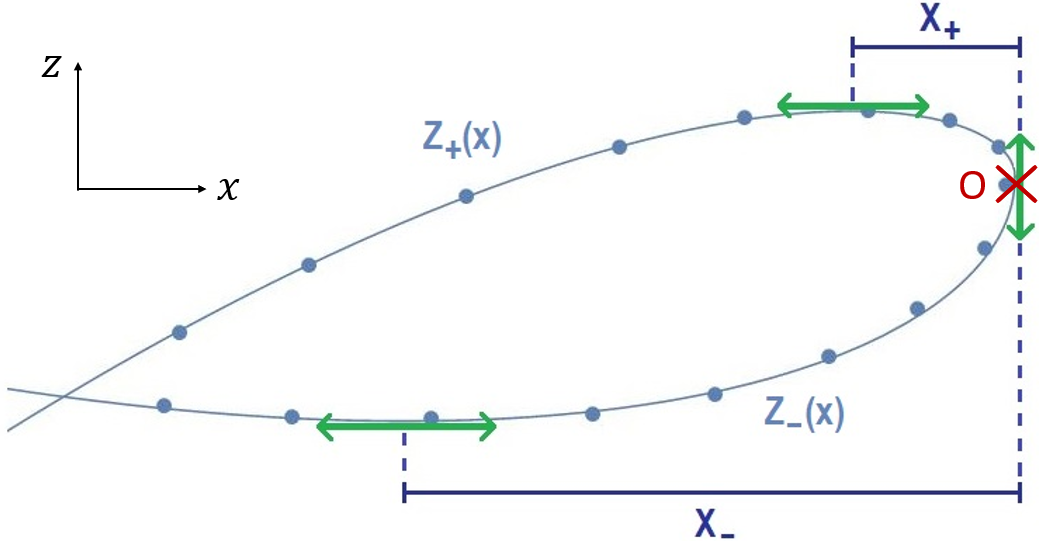}
    \caption{Illustration of the solutions $z_{\pm}$ and parameters $x_{\pm}$. The origin of the axes is at O.}
    \label{fig:x+-}
\end{figure}

The integration constants vanish if we appropriately choose the origin for $\vec{r}$ (i.e. where $x=z=0$) as the point O where the tangent to the loop is vertical, at the opposite side of the wheels (see fig~(\ref{fig:x+-})).

Taking the cross product of equation~(\ref{equa:vectoIntegre}) with $\vec{\tau}$, we obtain equation~(\ref{fig:integrodiff}) 
where $\theta$ is defined in the local Frenet-Serret frame (see fig~(\ref{fig:frenet})). Note that $\tan \theta = \frac{\upd z}{\upd x}.$

\begin{equation}
    -f \cdot (x \tan \theta - z) + \mu g s = 0
    \label{fig:integrodiff}
\end{equation}

This integro-differential equation can be solved analytically, and we get in equation~(\ref{equa:finale}) the following solutions $z_+ (x)$ and $z_- (x)$ respectively describing the upper and lower branches of the SSh's loop, 
where we define $x_{\pm}$ to be where $\frac{\upd z_{\pm}}{\upd x}$ vanishes (see fig~(\ref{fig:x+-})) and $\mathbf{R= \frac{\mu g}{f}}$ .
\begin{equation}
    \boxed{z_{\pm} (x) = \frac{1}{2} \bigg( \frac{x}{1 \pm R} \bigg| \frac{x}{x_{\pm}} \bigg|^{\pm R}  - \frac{x}{1 \mp R} \bigg| \frac{x}{x_{\pm}} \bigg|^{\mp R} \bigg)}
    \label{equa:finale}
\end{equation}

Because of the singularity when $x$ goes to $0$, we must impose that $R<1$ in order to have a finite mathematical solution. Physically, this means that the SSh should be able to remain in the air if and only if $R<1$, i.e. if and only if the air drag prevails over weight.

\subsection{Curve fitting for model-testing}

In order to test this model, we would like to experimentally vary the parameters that appear in our equations, i.e. the geometric parameters $x_+$ and $x_-$ (equation~(\ref{equa:finale})) as well as the physical parameter $R$.


The $x_+$ and $x_-$ parameters cannot be varied directly but they are determined by two other parameters controlled experimentally, namely the ejection angle  $\theta_0$ and the total length $L$ of the string. $R$ depends on the string's linear mass density $\mu$, but also its velocity $v$ and its features (such as the perimeter of a string section or its texture), which both influence the linear drag force applied on the string. The ambient medium in which the string is immersed can also be changed, as we did so qualitatively through the vacuum bell experiment.

For each configuration, we picked several points along the curve of the SSh. We have then used Mathematica \cite{b.mathematica} to compare the results of our model (equation~(\ref{equa:finale})) with these experimental points by curve fitting. This enabled us to get the values of the parameters $x_+$, $x_-$, $R$ and study the impact of the control parameters $v$, $\mu$, $L$ and $\theta_0$.

\section{Results}

\subsection{Two states and a takeoff speed}

\begin{figure*}[ht]
\begin{center}
    \begin{subfigure}{\columnwidth}
        \includegraphics[width=.9\columnwidth]{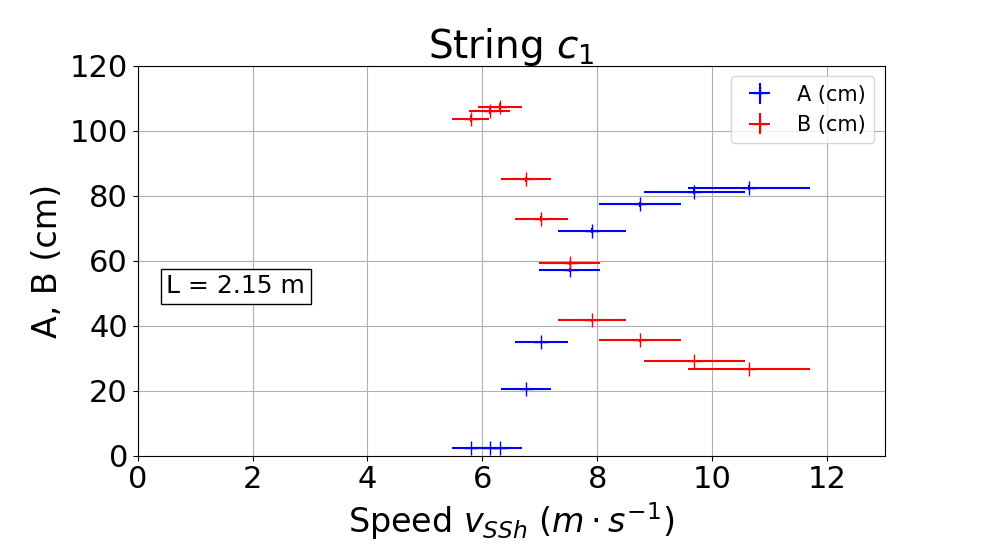}
    \end{subfigure}
    \begin{subfigure}{\columnwidth}
        \includegraphics[width=.9\columnwidth]{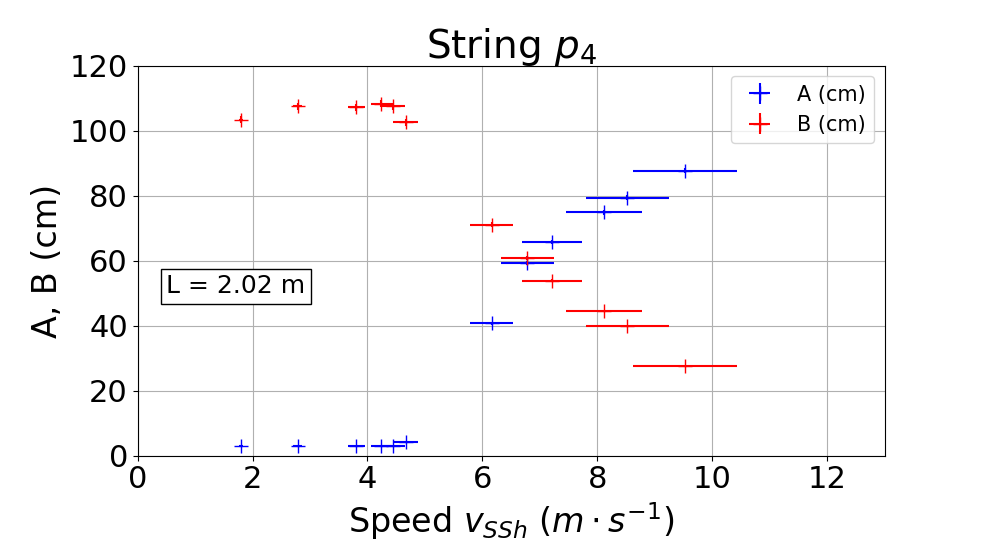}
    \end{subfigure}
    \begin{subfigure}{\columnwidth}
        \includegraphics[width=.9\columnwidth]{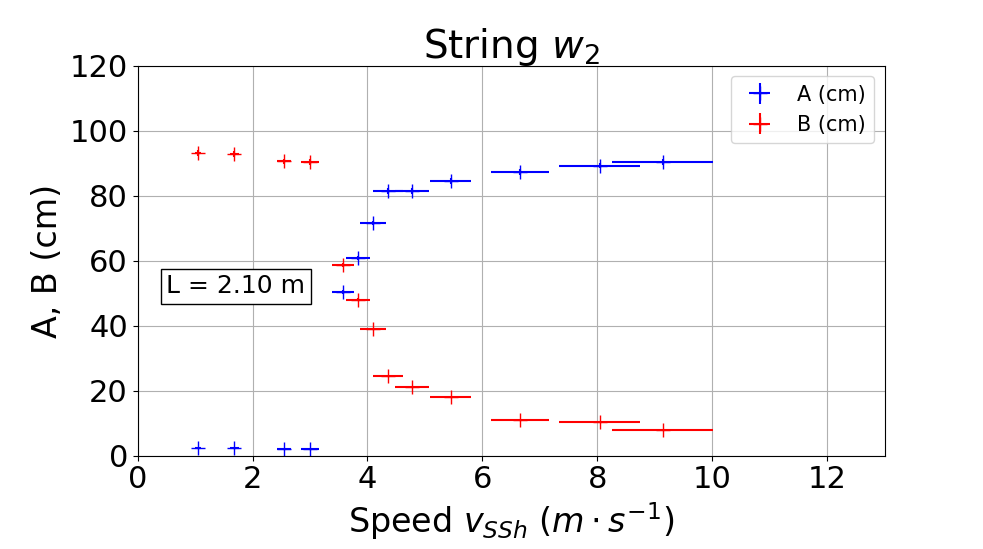}
    \end{subfigure}
    \begin{subfigure}{\columnwidth}
        \includegraphics[width=.9\columnwidth]{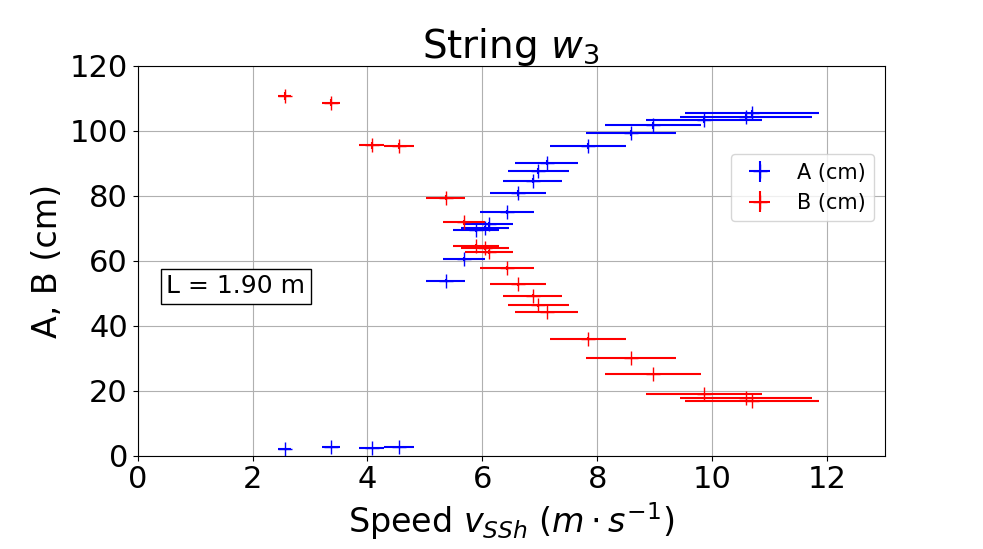}
    \end{subfigure}
    
    
    
    \caption{Evolution of the distances A and B as a function of the speed $v$, with $\theta_0 = 44.86 \degree$.}
    \label{fig:abV}
\end{center}
\end{figure*}

First of all, we studied for each string referenced in fig~(\ref{tab:strings}) how the SSh's speed $v$ influences the shape of the SSh by successively measuring the distances A and B as a function of $v$, see fig~(\ref{fig:abV}).

The graphs on fig~(\ref{fig:abV}) highlight the existence of two states separated by a particular threshold. Indeed, as we increase the speed of the SSh, it will at some point rise in the air. We decided to name the first phase the \emph{pulley state}, where the SSh still dangles from the shooter, and the second phase the \emph{air-lifted state}, where the loop starts rising in the air. As long as we remain in the pulley state, A and B remain constant. They start changing when the SSh is in the air-lifted state. We therefore define the takeoff speed $v_{\text{takeoff}}$ as the threshold between these two states and we notice that every string has its own and distinct takeoff speed (see table~(\ref{tab:seuils})).

In order to determine this takeoff speed experimentally, we calculate an average between the speed of the last point in the pulley state and the speed of the first point in the air-lifted state.

\begin{table}[ht]
\centering
\begin{tabular}{c c}
    \hline
     String used & $v_{\text{takeoff}}$ ($\un{m} \cdot \un{s^{-1}}$) \\
     \hline
     $c_1$ & $6.3 \pm 0.4 $ \\
     $w_2$ & $2.6 \pm 0.2$ \\
     $w_3$ & $4.6 \pm 0.3$ \\
     $p_4$ & $4.6 \pm 0.3$ \\
     $p_5$ & $5.6 \pm 0.5$ \\
     $p_6$ & $10.7 \pm 1.1$ \\
     \hline
     
\end{tabular}

\caption{Takeoff speed $v_{\text{takeoff}}$ for each of our strings.}
\label{tab:seuils}
\end{table}

\subsection{Dependence of $v_{\text{takeoff}}$}

Based on the results on table~(\ref{tab:seuils}), we noticed that the most dense strings as well as the smoothest strings were the ones to have the highest takeoff speeds $v_{\text{takeoff}}$. Besides the dependence on linear mass $\mu$, we wanted to determine whether $v_{\text{takeoff}}$ would also vary according to other characteristic parameters. Although these results are counterintuitive, it was found that for a given string, $v_{\text{takeoff}}$ depend neither on the length L of the SSh (see table~(\ref{table:indeL})) nor the angle $0 \degree < \theta_0 < 90 \degree$ (see  table~(\ref{table:indeTheta})).\footnote{The $v_{\text{takeoff}}$ on these tables are lower than those on table \ref{tab:seuils}: in the former experiments, the strings were slightly more worn out than in the latter experiments, which may have induced more air drag and thus facilitated its takeoff.}
Furthermore, the strings $w_3$ and $p_4$ have the same $v_{\text{takeoff}}$ despite having different linear mass $\mu$, which also proves that there must be other parameters specific to each string (such as its diameter, its material\ldots) that can influence their properties.

\subsection{The key role of air drag}

The linear air drag $f$ introduced in the equation of motion~(\ref{equa:Newton}) leads to a modified catenary equation. We therefore expect the overall shape of the SSh not to have a hyperbolic cosine profile as with an ordinary catenary, but rather to have a less common profile as described by equation~(\ref{equa:finale}).

\begin{figure*}[ht]
    \centering
    \includegraphics[width=\textwidth]{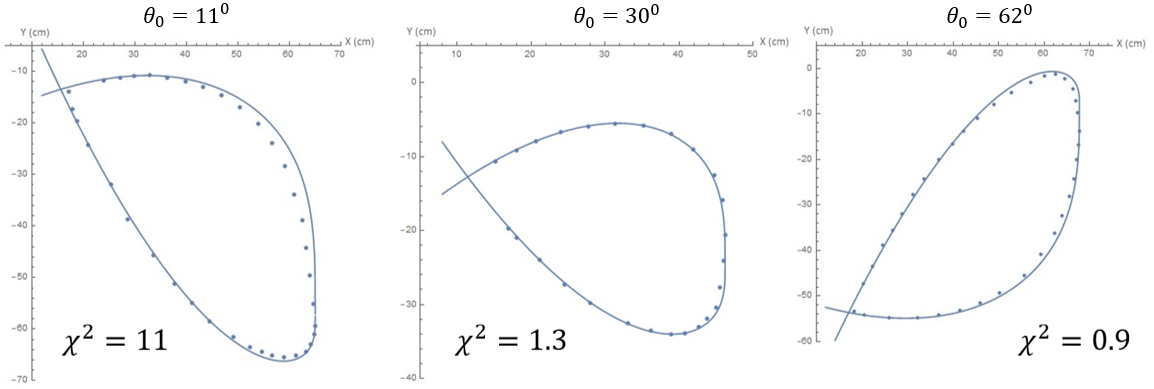}
    \caption{Curve fitting of our theoretical solution (eq.~(\ref{equa:finale})) with experimental profiles, with string $w_3$, $L = 160 \un{cm}$, $v = 5.1 \un{m}\cdot \un{s^{-1}}$ for several values of $\theta_0$.}
    \label{fig:fits}
\end{figure*}

The theoretical shape of the SSh, which was plotted with Mathematica by adjusting the $x_+$, $x_-$ and $R$ parameters, match our experimental measurements well, with $\chi^2 \leq 11$ (see fig~(\ref{fig:fits})). This provides a strong argument in favour of our model of the action of air on the string. 

    

\begin{table}[ht]
\centering
\begin{tabular}{c c}
    \hline
    \multicolumn{2}{c}{String $w_3$ at $\theta_0 = 20 \degree$} \\
    \hline
    Length $L$ ($\un{cm}$) & $v_{\text{takeoff}}$ ($\un{m}\cdot \un{s^{-1}}$) \\
    82 & $3.8 \pm 0.3 $ \\
    92 & $3.7 \pm 0.3$ \\
    145 & $3.6 \pm 0.3$ \\
    \hline
    \multicolumn{2}{c}{String $w_2$ at $\theta_0 = 20 \degree$} \\
    \hline
    Length $L$ ($\un{cm}$) & $v_{\text{takeoff}}$ ($\un{m}\cdot \un{s^{-1}}$) \\
    81 & $3.1 \pm 0.3 $ \\
    124 & $2.8 \pm 0.3$ \\
    161 & $2.9 \pm 0.3$ \\
    \hline
    
\end{tabular}
\caption{$v_{\text{takeoff}}$ in function of $L$ for strings $w_2$ and $w_3$.}
\label{table:indeL}
\end{table}

\begin{table}[ht]
\centering
\begin{tabular}{c c}
    \hline
    \multicolumn{2}{c}{String $w_3$ at $L = 82 \un{cm}$} \\
    \hline
    Initial angle $\theta_0$ ($\degree$) & $v_{\text{takeoff}}$ ($\un{m}\cdot \un{s^{-1}}$) \\
    20 & $3.8 \pm 0.3$ \\
    30 & $3.8 \pm 0.3$ \\
    40 & $4.1 \pm 0.3$ \\
    \hline
    \multicolumn{2}{c}{String $w_2$ at $L = 124 \un{cm}$} \\
    \hline
    Initial angle $\theta_0$ ($\degree$) & $v_{\text{takeoff}}$ ($\un{m}\cdot \un{s^{-1}}$) \\
    20 & $2.8 \pm 0.3 $ \\
    30 & $2.9 \pm 0.3$ \\
    40 & $2.8 \pm 0.3$ \\
    \hline
    
\end{tabular}
\caption{$v_{\text{takeoff}}$ in function of $\theta_0$ for strings $w_2$ and $w_3$.}
\label{table:indeTheta}
\end{table}

\section{Discussion}

\subsection{Analysis of the pulley and air-lifted states}

When $v < v_{\text{takeoff}}$, it behaves like a simple string that dangles down from a pulley (hence the name \emph{pulley state}). In this speed range, $v$ is not sufficient to maintain the SSh "in flight", as its weight dominates over air drag.
However, as soon the SSh reaches the threshold speed $v_{\text{takeoff}}$, air drag dominates over the weight and begins to lift the SSh.

Measurement uncertainties for $v_{\text{takeoff}}$ are unavoidable, such as those resulting from direct measurements of the length $L$ and the period $T_{\text{SSh}}$ of the SSh. The relative error on $L$ however is less than one percent, which makes it negligible (1cm of error out of 1m for $L$). The main source of uncertainties in our case stems from the measurement of $v$.

\subsection{Theoretical interpretation of $v_{\text{takeoff}}$}

Our theory naturally yields a dimensionless quantity $R = \mu g /f$ to describe the takeoff behaviour. $v_{\text{takeoff}}$ is only reached when $R=1$, i.e. when the weight and the air drag balance each other out. $R>1$ corresponds to the pulley state, while $R<1$ corresponds to the air-lifted state of the SSh, in which case we can mathematically describe it.

Note that this quantity $R$ does not depend on the length $L$ of the SSh, as both its total weight and total air drag are proportional to $L$.

\begin{figure}[ht]
    \centering
    \includegraphics[width=\columnwidth]{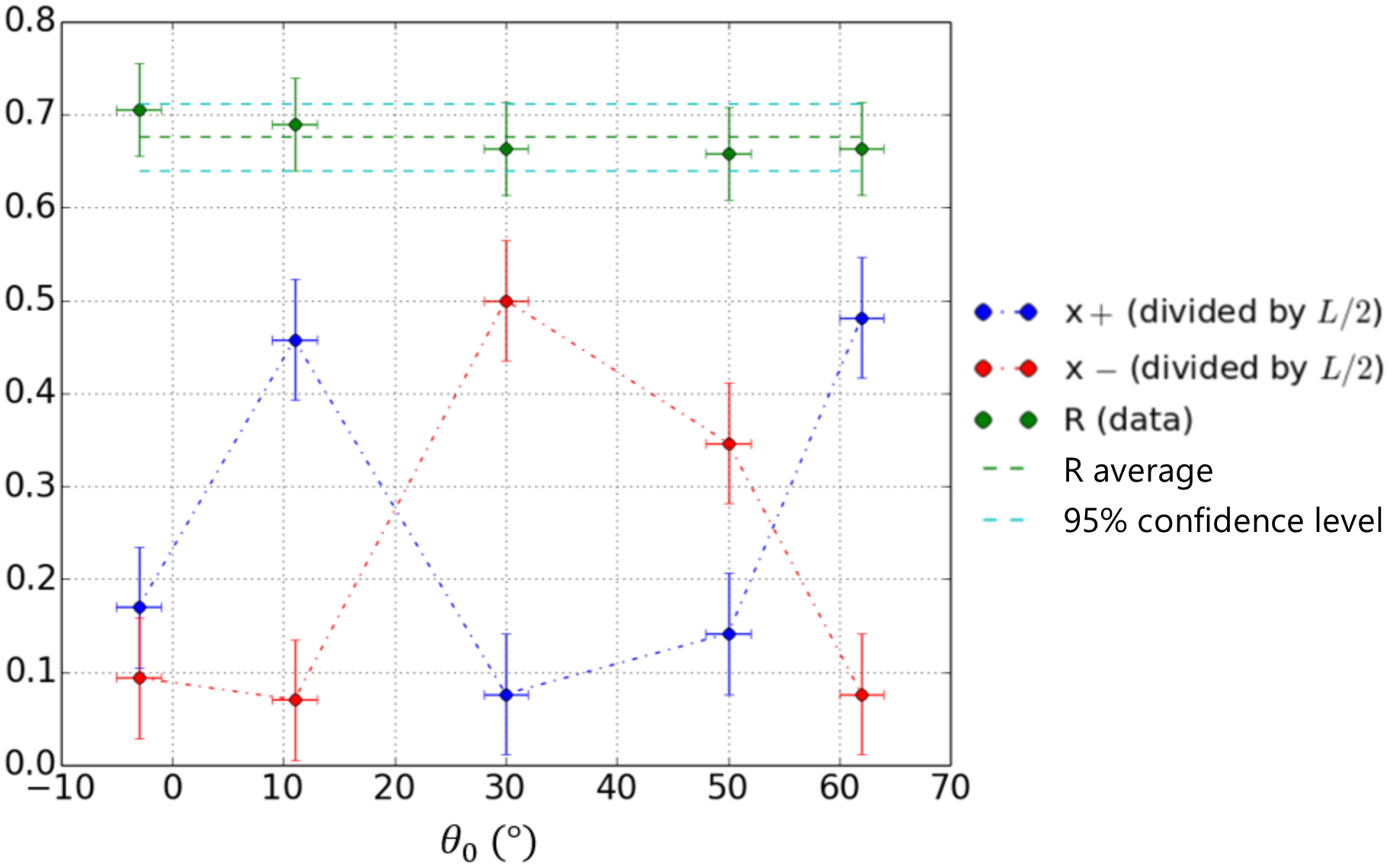}
    \caption{Evolution of the fit parameters $x_+$, $x_-$ and $R$ (with its average and 95\% confidence level) while varying $\theta_0$, keeping $v = 5.1 \un{m}\cdot \un{s^{-1}}$, $L = 160 \un{cm} $ and string $w_3$, in the air-lifted state ($R<1$).}
    \label{fig:Jul1}
\end{figure}

Likewise, the angle $\theta_0$ only intervenes on the boundary conditions: it does not directly appear in the equation of motion~(\ref{equa:Newton}). Therefore, the angle $\theta_0$ has no influence on the parameter R, and this is coherent with the experimental data (see fig~(\ref{fig:Jul1})). We observe the SSh's profile changes as $\theta_0$ varies, but in our mathematical formulation, this simply corresponds to a modification of the geometric parameters $x_+$ and $x_-$, which appear when integrating the equation of motion.

\begin{figure}[ht]
    \centering
    \includegraphics[width=\columnwidth]{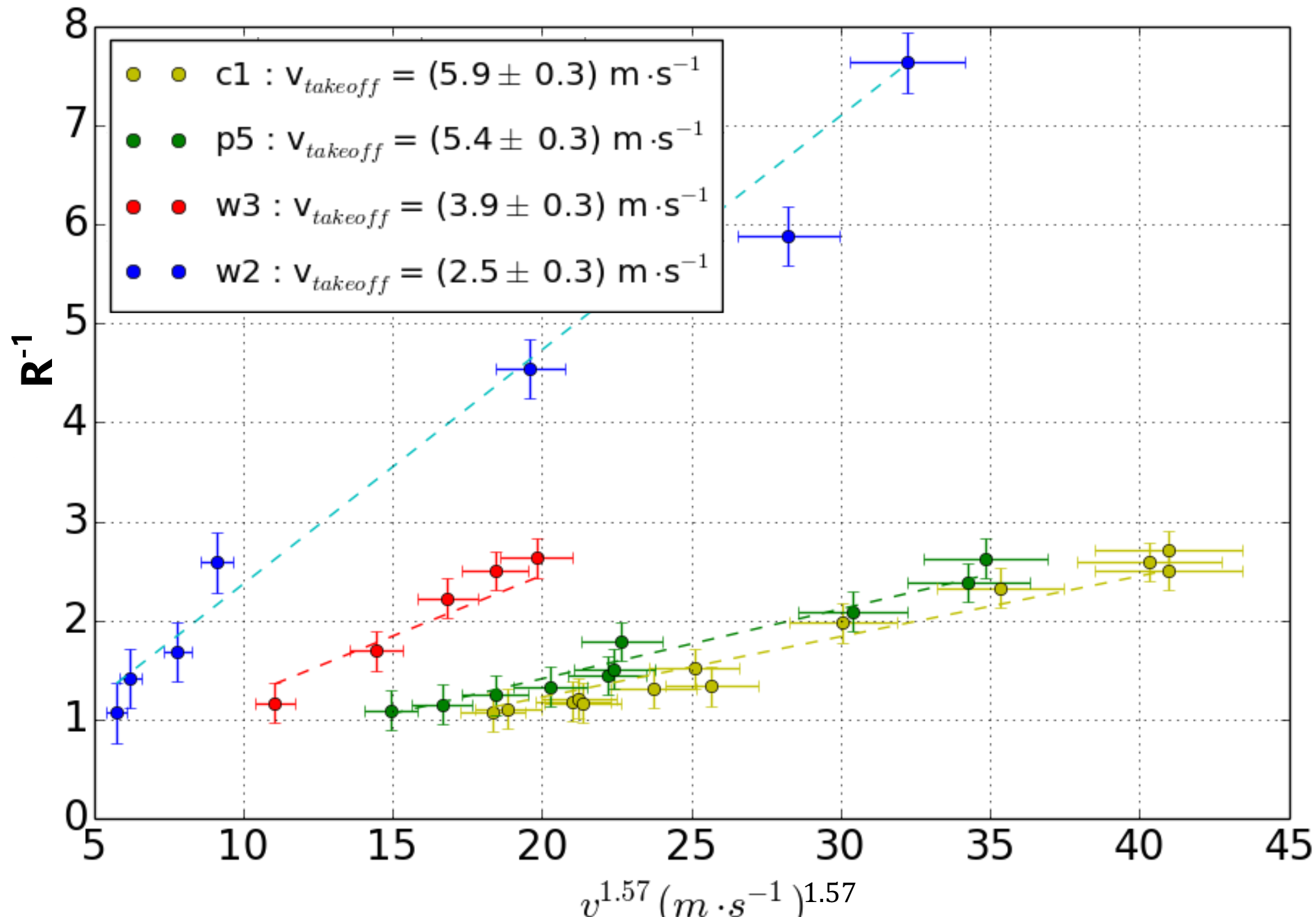}
    \caption{ $R^{-1} $ in function of $v^{1.57}$, with $R$ coming from curve fits based on several strings at different speeds in the air-lifted state ($R^{-1}>1$).}
    \label{fig:Jul2}
\end{figure}

Another point of interest is the dependence of the linear air drag $f$ on $v$. A common ansatz \cite{b.trainee} is to write $f$ as:
\begin{equation}
    f(v) = \alpha v^k
\end{equation}
with $\alpha > 0$ and $k \in [1 ; 2]$.

Note that since $R = \mu g /f$ and $1 = \mu g / f_{\text{takeoff}}$ (where $f_{\text{takeoff}}$ is the threshold linear air drag analogous to $v_{\text{takeoff}}$), we have:
\begin{equation}
    R = \bigg( \frac{v_{\text{takeoff}}}{v} \bigg) ^k.
\end{equation}

Using several linear fits and searching for $k$ with the lowest global $\chi^2$, we show that $1/R$ 
seems in fact proportional to $v^{1.57}$ (see fig~(\ref{fig:Jul2})). It means the air flow along the string is neither perfectly laminar $(k=1)$ nor totally turbulent $(k=2)$ but intermediate. This is consistent with the typical speed reached by our strings which is of the same order of magnitude as their respective takeoff speeds, themselves close to their terminal velocities in free fall, at which both air flow types are significant.

Furthermore, the linear coefficients that appear in fig~(\ref{fig:Jul2}) are related to the theoretical $v_{\text{takeoff}}$, 
which are in agreement with our experimental measurements 
(see table~(\ref{tab:seuils})). This further corroborates our model's validity.

\subsection{SSh shape’s instability}

Even if the SSh’s shape had been considered stable at each measurement, it appears that in practice, the SSh's shape sometimes takes a certain time to stabilise, particularly near $v_{\text{takeoff}}$. The irregularity of the shape could be coming from the loop's junction point passing through the sheath, or from the fact that the wheels' surface is imperfect (because of wear), as each of these issues could contribute to a possible systematic disruption of the system.

Because of this instability, we were careful that for every selected measurement, the loop would maintain a certain degree of stability: its shape should stay the same for at least a minute. Discrepancies relative to this stable shape were small enough that we could limit our systematic relative error on the distances A and B by 5\%.

\subsection{Qualitative results from a fan}

A qualitative study on the SSh in pulley mode allowed us to ensure that it is the air drag on the string that makes the SSh lift up in the air. The experiment consisted of putting a hand-held fan spinning at full speed\footnote{The air flow has a velocity of $4.5\un{m}\cdot \un{s^{-1}}$ at $10 \un{cm} $ and $2.9\un{m}\cdot \un{s^{-1}}$ at $30 \un{cm}$.} in front of the SSh's upper branch, so as to create a flux of air in the opposite direction for the SSh. We observed that the fan could make the string partially take off, even though we still had $v < v_{\text{takeoff}}$.

\subsection{Possible improvements}

\begin{figure}[ht]
    \centering
    \includegraphics[width=\columnwidth]{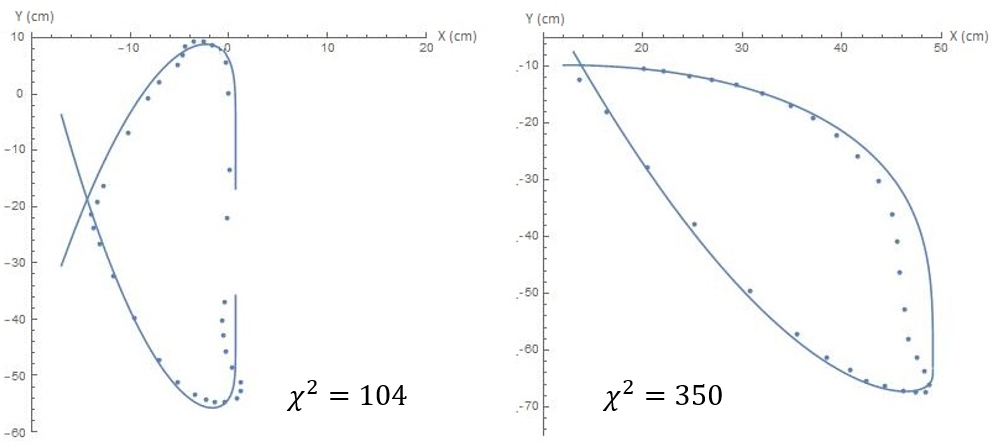}
    \caption{Examples of some cases where curve fitting using our theoretical model is not satisfactory (using string $w_2$).}
    \label{fig:fitsBof}
\end{figure}

We spent several months studying the SSh and further investigations will be required to fully understand its behaviour. As such, we believe it would be valuable and meaningful to explore the following leads:
\begin{itemize}
    \item working at lower and higher air pressure, or even changing the fluid in which the SSh lives, in order to further adjust and study the influence of drag;
    \item modifying the string itself by independently changing each of its characteristic parameters, such as linear density, roughness, fibre material, diameter, etc.;
    \item investigating the nature of the air drag and characterise the air flow along the string and therefore how $v$ influences the air drag, in order to calculate the value of k and understand if it changes in function of $v$;
    \item developing a more elaborate model accounting for the string's stiffness or aerodynamics beyond just air drag, such as the Magnus effect, to more precisely describe deviations from the general profile predicted by our theory: fig~(\ref{fig:fitsBof}) indeed shows that the curve predicted by the model is only valid for the general shape of the SSh, but incorrectly describe its shape at the opposite side of the motor;
    \item plotting an average profile of SSh taken from separate measurements. For this, we would need elaborate an automatic image processing program, as hundreds of measurements would be analysed for each configuration.

\end{itemize}



\section{Dead end}

\subsection{Experimental issues with the SSh setup} 

The design and manufacture of a SSh is not as simple as it may seem. Several  conceptual ideas turned out to be flawed. With speed $v$ in mind, we first tried using motors which had a very high no-load rotation speed but very little torque when loaded, which made them unusable for a SSh. We also tried using 3D-printed plastic wheels (covered with felt to improve grip), but there was still too much slip between the string and the wheels.

In the course of our experiments, we had recurring issues such as faulty wheel alignments or the string entangling itself in the wheels. The most severe issue however was wear, especially at higher speeds (our rubber wheels, or "tyres", would considerably deform elastically at these speeds). Higher speeds meant more mechanical stress on every component of the SSh (string, wheels, sheath). 

This shows that our quest for high speeds was troublesome, and simply swapping our existing motor for a faster and more powerful one would put too much stress on our current SSh to bear. More sturdy and reliable components along with extensive engineering work beyond our means would have been required to reach much higher speeds. Our SSh thus had a practical "speed limit", which meant that we were unable to reach the takeoff speed for some of our strings.

\subsection{Ballasted string drop}

Since the air drag is quadratic, i.e. $f= \alpha v^2 $ with $\alpha$ linked to the properties of a given string, then at the threshold $R=1$, we would get $v_{\text{takeoff}} = \sqrt{ \mu g / \alpha} $. Furthermore, a falling object subject to such a drag force has a terminal velocity of $v_{\text{terminal}} = \sqrt{ \mu g / \alpha} $. We have therefore attempted to measure this terminal velocity in order to experimentally compare it to the \emph{takeoff speed}.

A sample of string of length around $l=50\un{cm}$ attached to a metal nail was dropped without initial velocity from a height of around $20 \un{m}$. Its speed was measured using video footage taken just above the ground. The ballast ensures that the string sample is aligned with its downwards movement, like a cylinder along its main axis (like in the \emph{SSh}'s stationary state). As the terminal velocity results from an equilibrium between weight and air drag, it is possible to isolate the additional mass term of the nail to eliminate it.

Unfortunately, this experiment was not conclusive. For most of our types of strings, either the terminal velocity was not reached early enough, or the string did not keep a cylindrical and longitudinal shape along its movement. Additionally, we could not guarantee that the influence of nail aerodynamics on the string was negligible.

\subsection{Wind tunnel experiment}

In order to confront \emph{takeoff speed} and \emph{terminal velocity} for a given string material, we also designed an experiment using a wind tunnel to try to determine the drag coefficient $\alpha$. The main idea was to use the wind tunnel in a \textbf{vertical} configuration. A string sample would be glued to a very thin inelastic string: the sample would dangle in the wind tunnel while the thin string would be attached to a precise dynamometer\footnote{A resolution of $10^{-5}$ N was reachable with our equipment.} outside the tunnel. While varying the turbine’s power, we would measure both the wind velocity $v$ and the drag force exerted on the string sample (after subtracting its weight), thereby obtaining $\alpha$ by regression. This would have then provided an accurate value for the \emph{terminal velocity}, which we could have compared to the \emph{takeoff speed} seen on the SSh.
 
Unfortunately, while the rest of the experimental setup was implemented as envisioned, the wind tunnel at our disposal was in a \textbf{horizontal} configuration. Working in a horizontal configuration was not possible, as we wished to study the drag in the case where the string sample is aligned with the wind: the string sample therefore had to be collinear with its weight. Given the dimensions of the tunnel and the weight of the turbine, adapting it to a safe and stable vertical configuration would have required extensive engineering resources specifically tailored to our needs, which were not available to us.




\section{Conclusion}

Having discovered that air drag plays a significant and crucial role in the physics of the SSh, we have explained how there are two possible distinct states for the SSh: the \emph{pulley state} where the loop dangles down, and the \emph{air-lifted state}. We have experimentally shown that the transition between these states depends on the nature of the string rather than on the loop’s length $L$ or its angle of launch $\theta_0$.

Using a mathematical model, we have shown that a drag term is indeed necessary to describe an \emph{air-lifted} state and that there indeed exists a threshold between these two states when the air drag equals the weight. The validity of this model in describing the overall shape of the SSh was proven by curve fitting on experimental profiles. Furthermore, the quadratic nature of the air drag was proven simply by comparing the dimensionless quantity $R$ (obtained by curve fitting) to the SSh's speed $v$.

After the submission of this paper, we have become aware of two papers recently published on the same topic \cite{b.cambridge,b.physrev}, which put more emphasis on the study of wave propagation along the string. We were also informed of the existence of \cite{b.journalmech} which was the first to theorise the geometrical shape of the SSh in 1949.




\acknowledgments
This study could not have been completed without the generous support of our supervisors Frédéric Bouquet, Frédéric Chevy, Frédéric Moisy, Adrian Daerr and Nabil Garroum throughout this project, without forgetting our colleagues Clément Abadie, Jeanne de Courson, Usama Iqbal and Pierre Terpereau. We thank them greatly.

We especially wish to thank our laboratory technician Wladimir Toutain for his invaluable help in the manufacture of the String Shooter setup.


\begin{thebibliography}{0}

\bibitem{b.yt1}
  \Name{Yeany B.}
  "String shooter-String launcher- physics of toys - Homemade Science with Bruce Yeany",
  YouTube video, 4:19,
  June 11, 2014, https://youtu.be/rffAjZPmkuU
  .

\bibitem{b.yt2}
  \Name{d'Art of Science}
  "How to make a Shooter - String Shooter",
  YouTube video, 6:54,
  June 28, 2018, https://youtu.be/bsZ93j\_nUVc
  .

\bibitem{b.yt3}
  \Name{CJ Gilbert}
  "Science \& Technology Podcast, Episode 33: String Shooter",
  YouTube video, 7:58,
  Januuary 10, 2017, https://youtu.be/Tcp0lgQOT6c
  .

\bibitem{b.yt4}
  \Name{Scorch Works}
  "String Shooter",
  YouTube video, 6:03,
  July 4, 2018, https://youtu.be/IGIDy-IJxmo
  .

\bibitem{b.curv}
  \Name{Salençon J.}
  \Book{Mécanique des milieux continus, Tome III, Milieux curvilignes}
  \Editor{Éditions de l'École Polytechnique}
  \Year{2016}.

\bibitem{b.audacity}
    \Book{Audacity (version 2.3.0). Windows.}
    \Year{Downloaded on 24/10/2018 at https://www.audacityteam.org/download/}.

\bibitem{b.ImageJ}
    \Book{ImageJ (version 1.52). Windows.}
    \Year{Downloaded on 19/03/2019 at https://imagej.nih.gov}.

\bibitem{b.IC}
    \Book{IC Measure (version 2.0.0.161). Windows. The Imaging Source.}
    \Year{Downloaded on 23/01/2019 at https://www.theimagingsource.com}.

\bibitem{b.mathematica}
    \Book{Mathematica (version 12.0). Windows. Wolfram Research.}
    \Year{Downloaded on 04/05/2019 at https://www.wolfram.com/mathematica/}.

\bibitem{b.webplot}
    \Name{Rohatgi A.}
    \Book{WebPlotDigitizer (version 3.9). Windows.}
    \Year{Downloaded on 17/01/2019 at https://automeris.io/WebPlotDigitizer/index.html}.
    
\bibitem{b.trainee}
    \Name{Candel S.}
    \Book{Mécanique des fluides}
    \Editor{Dunod, Paris}
    \Year{1990}
    \Pages{305}{310}.

\bibitem{b.cambridge}
    \Name{Daerr A., Courson J., Abello M., Toutain W. \& Andreotti B.}
    "The charmed string: self-supporting loops through air drag",
    \Book{Journal of Fluid Mechanics}
    \Vol{877},
    \Publ{Cambridge University Press},
    \Year{2019}.

\bibitem{b.physrev}
    \Name{Taberlet N. et al.}
    "Propelled Strings: Rising from Friction",
    \Book{Phys. Rev. Lett.}
    \Vol{123},
    \Publ{American Physical Society},
    \Year{2019}.
    
\bibitem{b.journalmech}
    \Name{Gregory C.C.L.}
    "Theory of a loop revolving in air, with observations on the skin-friction",
    \Book{The Quarterly Journal of Mechanics and Applied Mathematics}
    \Vol{2},
    \Publ{Oxford Academic},
    \Year{1949}.




\end{thebibliography}
\end{document}